# Rise time of spherical intruder in granular fluid

Sparisoma Viridi[a], Nuning Nuraini[b], Mohammad Samy[c], Ayu Fitriyanti[c],
Ika Kusuma Adriani[c], Nurwenda Amini[c], and Ganjar Santoso[c]

May 16, 2011 v1

[a]Nuclear Physics and Biophysics Research Division
[b]Industrial and Financial Mathematics Research Division
[c]Mathematics Study Program
Institut Teknologi Bandung, Jalan Ganesa 10, Bandung 40132, Indonesia
[a]dudung@fi.itb.ac.id

Abstract

A model is developed to explain rise time of a spherical intruder placed in a granular bed, which is considered as fluid. Phenomenon of rising intruder in a granular bed is well known as Brazil nut effect. Radius of the intruder is varied with $R_n = nR_0$, $n = 1..10$. An approximation for $t \ll \tau$ is chosen in order to simplify the solution of second order differential equation of intruder vertical position to obtain the rise time $T$. A non-physical parameter $\alpha$ and also transformation from $T$ to $T'$ are needed to be introduced in order to control the results to mimic reported experiment qualitatively. Several forms of rise time $T$ and also transformed rise time $T'$ against $n$ are presented and discussed.

## I. Introduction

A work via molecular dynamics (MD) simulation for one and several intruders showed the phenomenon known as Brazil nut effect (BNE) [1]. Particles density between intruder and the granular bed plays important role in size separation of the grains [2], even a large particle could sink instead of rising [3]. Rise time of BNE and its reverse (RBNE) have been explained through hydrodynamic theory based on the Boltzmann-Enskog theory [4]. In this work a simple approach by considering a vibrated granular bed as fluid is chosen. Since it is considered as fluid it should have viscosity $\eta$ and introduce a drag force $F_d$ to the intruder as it is rising. The form of drag force is the same form as the known form in Stoke's law [5]. Other forces considered are gravitational force $F_g$ and buoyant force $F_a$. Applying Newton's second law of motion will lead to a second order differential equation whose solution is then used to estimate the rise time $T$.

## II. Theoretical model

An spherical intruder with diameter $R_n$ will be used, with

$$R_n = nR_0, \quad n = 1..10. \tag{1}$$

Vertical direction is set to be in $y$-direction where gravitation acceleration $g$ is taken to be in negative $y$-direction. Then the gravitational force $F_g$ suffered by the intruder with density $\rho_b$ is

$$F_g = -\rho_b g V_n. \tag{2}$$

with $V_n$ is immersed volume of the spherical intruder

$$V_n = \frac{4}{3}\pi R_n^3, \quad n = 1..10. \tag{3}$$

Since the granular bed is considered as fluid, with density $\rho_f$ and viscosity $\eta_f$, than it will give a buoyant force $F_a$ and drag force $F_d$ to the intruder, which are

$$F_a = \rho_f g V_n, \qquad (4)$$

$$F_d = -6\eta_f R_n \frac{dy}{dt}, \qquad (5)$$

Then by using Newton's second law of motion and Equation (2), (4) and (5), it can be obtained that

$$\frac{d^2 y}{dt^2} + \left(\frac{6\eta_f R_n}{m}\right)\frac{dy}{dt} = \left[\frac{(\rho_f - \rho_b)g V_n}{m}\right]. \qquad (6)$$

Equation (6) will be solved to obtained $y(t)$ which is then used to estimate rise time $T$. First, the right side of Equation (6) is set to zero and the equation can be written as

$$\frac{d}{dt}\left[\frac{dy}{dt} + \left(\frac{6\eta_f R_n}{m}\right)y\right] = 0. \qquad (7)$$

In general, Equation (7) will give condition that there is a constant $k$

$$\frac{dy}{dt} + \left(\frac{6\eta_f R_n}{m}\right)y = k. \qquad (8)$$

A new variable $y'$ can be introduced to simplify Equation (8), which is

$$y' = y - k\left(\frac{m}{6\eta_f R_n}\right), \qquad (9)$$

so that Equation (8) can be rewritten as

$$\frac{dy'}{dt} + \left(\frac{6\eta_f R_n}{m}\right)y' = 0. \qquad (10)$$

Solution of Equation (10) then will be

$$y' = A e^{-\left(\frac{6\eta_f R_n}{m}\right)t}. \qquad (11)$$

Equation (11) can be used with help of Equation (9) to get $y$, which is

$$y = A e^{-\left(\frac{6\eta_f R_n}{m}\right)t} + k\left(\frac{m}{6\eta_f R_n}\right), \qquad (12)$$

where there are two unknown constants $A$ and $k$, which will be determined. Equation (12) is solution of Equation (8), but it is not yet solution of Equation (6). In order to get solution of Equation (6), Equation (12) must be modified into

$$y = A e^{-\left(\frac{6\eta_f R_n}{m}\right)t} + Bt + k\left(\frac{m}{6\eta_f R_n}\right). \qquad (13)$$

Now, it is the time to determine the two unknown constants $A$, $B$, and $k$ by applying the boundary condition that at $t = 0$

$$y(0) = y_0 \text{ and } \frac{dy(0)}{dt} = v_0. \tag{14}$$

Then, it can be obtained that

$$y_0 = A + k\left(\frac{m}{6\eta_f R_n}\right) \tag{15}$$

and

$$v_0 = -\left(\frac{6\eta_f R_n}{m}\right)A + B \tag{16}$$

From Equation (15) and (16) the two unknown constants $A$ and $k$ can be found, which are

$$A = -\left(\frac{m}{6\eta_f R_n}\right)(v_0 - B) \tag{17}$$

and

$$k = \left(\frac{6\eta_f R_n}{m}\right)y_0 + (v_0 - B) \tag{18}$$

Using Equation (17) and (18), Equation (13) can be rewritten as

$$y(t) = \left[-\left(\frac{m}{6\eta_f R_n}\right)(v_0 - B)\right]e^{-\left(\frac{6\eta_f R_n}{m}\right)t} + Bt + \left[y_0 + \left(\frac{m}{6\eta_f R_n}\right)(v_0 - B)\right]. \tag{19}$$

The constant $B$ can be found using following equations

$$\frac{dy}{dt} = (v_0 - B)e^{-\left(\frac{6\eta_f R_n}{m}\right)t} + B, \tag{20}$$

$$\frac{d^2y}{dt^2} = -\left(\frac{6\eta_f R_n}{m}\right)(v_0 - B)e^{-\left(\frac{6\eta_f R_n}{m}\right)t}, \tag{21}$$

and substitution of Equation (19)-(21) into Equation (6) will give

$$-\left(\frac{6\eta_f R_n}{m}\right)(v_0 - B)e^{-\left(\frac{6\eta_f R_n}{m}\right)t} + \left(\frac{6\eta_f R_n}{m}\right)\left[(v_0 - B)e^{-\left(\frac{6\eta_f R_n}{m}\right)t} + B\right] = \left[\frac{(\rho_f - \rho_b)gV_n}{m}\right] \tag{22}$$

and then

$$B = \frac{(\rho_f - \rho_b)gV_n}{6\eta_f R_n}. \tag{23}$$

So, finally Equation (19) can be written as

$$y(t) = \left(\frac{m}{6\eta_f R_n}\right)\left[\frac{(\rho_f - \rho_b)gV_n}{6\eta_f R_n} - v_0\right]e^{-\left(\frac{6\eta_f R_n}{m}\right)t} + \left[\frac{(\rho_f - \rho_b)gV_n}{6\eta_f R_n}\right]t \\ + \left\{y_0 + \left(\frac{m}{6\eta_f R_n}\right)\left[v_0 - \frac{(\rho_f - \rho_b)gV_n}{6\eta_f R_n}\right]\right\}. \quad (24)$$

Equation (24) will be used to estimate the rise time $T$. Actually $T$ can be found by giving the condition when the intruder arrives at the top of granular bed or $y(T) = y_T$ and then solve Equation (24) using some method, e.q. self consistent field (SCF) method which is common for complex functions in quantum chemistry [6]. But in this work only simple approximation of Equation (24) considered to calculate estimate the rise time $T$.

## III. Approximation to calculate the rise time

It is assumed that there is a time $\tau$ that so small so that Equation (24) can be approximated into

$$y(t) \approx \left(\frac{m}{6\eta_f R_n}\right)\left[\frac{(\rho_f - \rho_b)gV_n}{6\eta_f R_n} - v_0\right]e^{-\left(\frac{6\eta_f R_n}{m}\right)t} + \left\{y_0 + \left(\frac{m}{6\eta_f R_n}\right)\left[v_0 - \frac{(\rho_f - \rho_b)gV_n}{6\eta_f R_n}\right]\right\} \quad (25)$$

for $t \ll \tau$ and, of course, for $t \gg \tau$ it can be obtained that

$$y(t) \approx \left[\frac{(\rho_f - \rho_b)gV_n}{6\eta_f R_n}\right]t + \left\{y_0 + \left(\frac{m}{6\eta_f R_n}\right)\left[v_0 - \frac{(\rho_f - \rho_b)gV_n}{6\eta_f R_n}\right]\right\}. \quad (26)$$

In this work oleh Equation (25) is considered. By setting that $y(T) = y_T$ it can obtained that Equation (25) will give

$$T = -\left(\frac{m}{6\eta_f R_n}\right)\ln\left[1 + \frac{(6\eta_f R_n)^2(y_T - y_0)}{(\rho_f - \rho_b)mgV_n - (6\eta_f R_n)mv_0}\right]. \quad (27)$$

Subtitue Equation (1) and (3) into Equation (27) will lead to

$$T_n = -\left(\frac{m}{6\eta_f R_0}\right)\frac{1}{n}\ln\left[1 + \frac{3(6\eta_f R_0)^2(y_T - y_0)n^2}{4(\rho_f - \rho_b)mg\pi R_0^3 n^3 - 3(6\eta_f R_0)mv_0 n}\right]. \quad (28)$$

Equation (28) can be simplified into

$$T_n = \frac{c_1}{n}\ln\left[1 + \frac{c_2 n^2}{c_3 n^3 + c_4 n}\right] \quad (29)$$

by defined the following constants

$$c_1 = -\left(\frac{m}{6\eta_f R_0}\right), \quad (30)$$

$$c_2 = 3(6\eta_f R_0)^2(y_T - y_0), \quad (31)$$

$$c_3 = 4(\rho_f - \rho_b)mg\pi R_0^3, \quad (32)$$

$$c_4 = -3(6\eta_f R_0)mv_0. \tag{33}$$

Equation (29) is the equation to be discussed in the next section.

## IV. Results and discussions

From the derivation in the previous section it can illustrated in the following figures for Equation (29). In varying value of $c_1$, $c_2$, $c_3$, and $c_4$, according to value of $m$, $\eta_f$, $R_0$, $y_T$, $y_0$, $\rho_f$, $\rho_0$, $g$, and $v_0$, it must be that $c_1 < 0$, $c_2 > 0$, $c_3 > 0$, and $c_4 = 0$. The latest value of parameter is because the intruder is considered at rest at $t = 0$. Then Equation (29) will reduced to

$$T_n = \frac{c_1}{n}\ln\left[1+\frac{c_5}{n}\right], \tag{34}$$

with

$$c_5 = \frac{c_2}{c_3}. \tag{35}$$

Since $c_2 > 0$ and $c_3 > 0$ then $c_5 > 0$. Unfortunately these constraints can not produce desired results as reported [7].

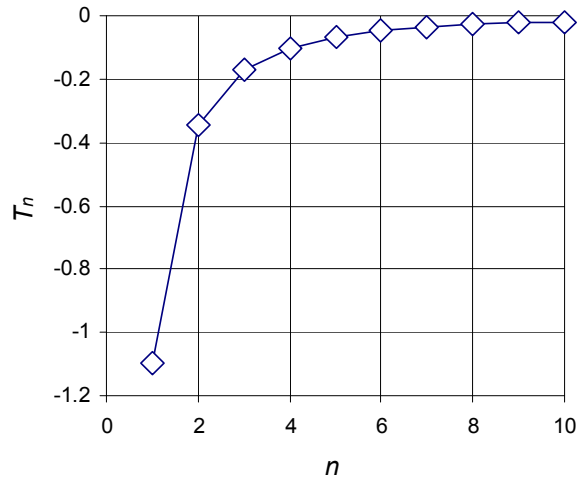

Figure 1. Illustration of Equation (34) with $c_1 = -1$ and $c_5 = 2$.

A modification is made by introducing a parameter $\alpha$ which is placed as in the following equation

$$T_n = \frac{c_1}{n^\alpha}\ln\left[1+\frac{c_5}{n}\right] \tag{36}$$

and the illustration with different value of $\alpha$ is given in Figure 2. This modification gives also better results than the previous equation. The reported experiment [7] shows that the profile of $T_n$ against $n$ is similar to an u-shape. But, that is interesting, when value of $\alpha$ is chosen around –0.88 the u-like-shape can be obtained as illustrated in Figure 3. Three values of $\alpha$ are given in Figure 3 to show the influence of the parameter in altering shape of curve of $T_n$ against $n$.

If Equation (36) is right, then there is still a major problem about value of $T_n$ which is negative. There has been yet any explanation why it can have value less than zero.

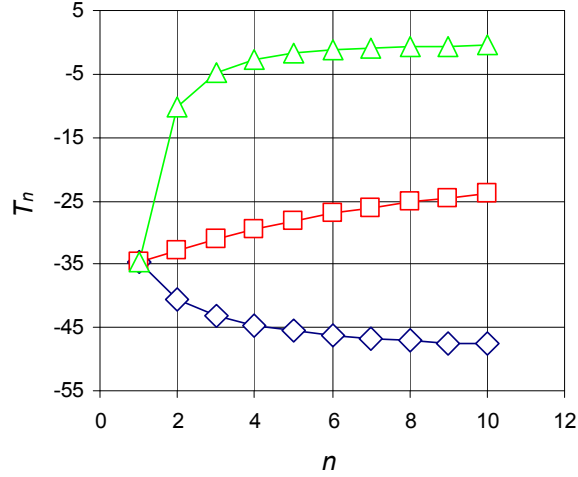

Figure 2. Illustration of Equation (36) with $c_1 = -50$ and $c_5 = 1$ for: $\alpha = -1$ (◊), $\alpha = -0.7$ (□), and $\alpha = 1$ (Δ).

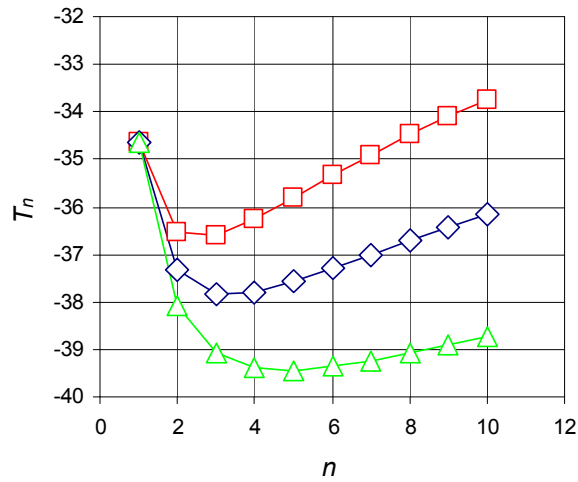

Figure 3. Illustration of Equation (36) with $c_1 = -50$ and $c_5 = 1$ for: $\alpha = -0.85$ (□), $\alpha = -0.88$ (◊), and $\alpha = -0.91$ (Δ), show a nearly u-shape profile.

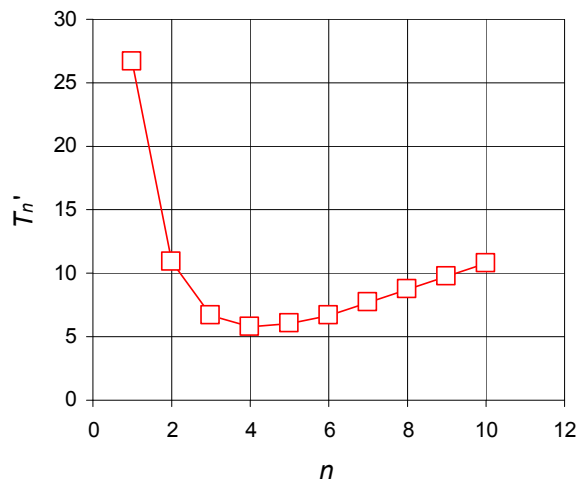

Figure 4. Results of Equation (36) and (37) for $c_1 = -50$, $c_5 = 1$, and $\alpha = -0.9$ shows a u-like-shape.

A transformation of this negative $T_n$ can also be done, e.g. using following equation

$$T_n' = 5(T_n + 40), \tag{37}$$

but loss of physical meaning, which gives Figure 4. Then it can be written that equation

$$T_n' = \frac{5c_1}{n^\alpha} \ln\left[1 + \frac{c_5}{n}\right] + 200, \tag{38}$$

with $c_1 = -50$, $c_5 = 1$, and $\alpha = -0.9$ will give a similar result to the reported experiment [7] but without fully physical background. Further modifications with stronger reasons are needed.

## V. Conclusions

It can be concluded that the u-like-shape profile of $T_n'$ against $n$ can be obtained. Unfortunately there are two unphysical reasons must be introduced. First is the parameter $\alpha$ and second is transformation from $T$ to $T'$. Further investigation is a must to explain the phenomenon or the two unphysical reasons.

## Acknowledgements

This work is supported partially by Institut Teknologi Bandung Alumni Research Grant in year 2009 and 2010.